\documentclass[12pt,manuscript]{aastex}

\usepackage{epsfig} \usepackage{natbib} 
\slugcomment{Draft - not for public release.}

\shorttitle{Lower bounds on photometric redshift estimation}
\shortauthors{Asztalos et al.}

\begin{document}

%% LaTeX will automatically break titles if they run longer than one
%% line. However, you may use \\ to force a line break if you desire.

\title{Lower bounds on photometric redshift errors \\from Type Ia
  supernovae templates}

%% Use \author, \affil, and the \and command to format author and
%% affiliation information.  Note that \email has replaced the old
%% \authoremail command from AASTeX v4.0. You can use \email to mark
%% an email address anywhere in the paper, not just in the front
%% matter.  As in the title, use \\ to force line breaks.

\author{S. Asztalos} \affil{XIA, LLC, Hayward, CA 94554}
\newcommand{\myemail}{SJAsztalos@lbl.gov}

\author{S. Nikolaev, W. de Vries, S. Olivier and K. Cook}
\affil{Lawrence Livermore National Laboratory, Livermore, CA 94551}

\author{L. Wang} \affil{Texas A\&M, College Station, TX 77843}

%% are identified by the \altaffilmark after each name.  Specify
%% alternate affiliation information with \altaffiltext, with one
%% command per each affiliation.

%% Mark off your abstract in the ``abstract'' environment. In the
%% manuscript style, abstract will output a Received/Accepted line
%% after the title and affiliation information. No date will appear
%% since the author does not have this information. The dates will be
%% filled in by the editorial office after submission.

\begin{abstract}
  Cosmology with Type Ia supernovae heretofore has required extensive
  spectroscopic follow-up to establish a redshift. Though tolerable at
  the present discovery rate, the next generation of ground-based
  all-sky survey instruments will render this approach unsustainable.
  Photometry-based redshift determination is a viable alternative, but
  introduces non-negligible errors that ultimately degrade the ability
  to discriminate between competing cosmologies. We present a strictly
  template-based photometric redshift estimator and compute redshift
  reconstruction errors in the presence of photometry and statistical
  errors. With reasonable assumptions for a cadence and supernovae
  distribution, these redshift errors are combined with systematic
  errors and propagated using the Fisher matrix formalism to derive
  lower bounds on the joint errors in $\Omega_w$ and $\Omega_{w'}$
  relevant to the next generation of ground-based all-sky survey.
\end{abstract}

%% Keywords should appear after the \end{abstract} command. The
%% uncommented example has been keyed in ApJ style. See the
%% instructions to authors for the journal to which you are submitting
%% your paper to determine what keyword punctuation is appropriate.


%% From the front matter, we move on to the body of the paper.  In the
%% first two sections, notice the use of the natbib \citep and \citet
%% commands to identify citations.  The citations are tied to the
%% reference list via symbolic KEYs. The KEY corresponds to the KEY in
%% the \bibitem in the reference list below. We have chosen the first
%% three characters of the first author's name plus the last two
%% numeral of the year of publication as our KEY for each reference.

%% linked in the electronic edition to a data center may do so by
%% tagging their objects with \objectname{} or \object{}.  Each macro
%% takes the object name as its required argument. The optional,
%% square-bracket argument should be used in cases where the data
%% center identification differs from what is to be printed in the
%% paper.  The text appearing in curly braces is what will appear in
%% print in the published paper.  If the object name is recognized by
%% the data centers, it will be linked in the electronic edition to
%% the object data available at the data centers
%%
%% Note that for sources with brackets in their names, e.g. [WEG2004]
%% 14h-090, the brackets must be escaped with backslashes when used in
%% the first square-bracket argument, for instance,
%% \object[\[WEG2004\] 14h-090]{90}).  Otherwise, LaTeX will issue an
%% error.

\section{Introduction}

Accumulating evidence for an accelerating universe
\citep{1998AJ....116.1009R,1999ApJ...517..565P,2006AA...447...31A,
  2007ApJ...666..694W} is premised on the photometry of several
hundred Type Ia supernovae whose redshifts were exquisitely determined
through lengthy spectroscopic follow-up. At the current supernovae
Type Ia discovery rate this situation remains manageable and
spectroscopic follow-up continues to be the norm. Nonetheless, we are
approaching an era of ground-based survey
telescopes\footnote{http://msowww.anu.edu.au/skymapper/}\footnote{http://pan-starrs.ifa.hawaii.edu/public/}\footnote{http://www.lsst.org/lsst}\footnote{http://www.darkenergysurvey.org/}
which may entail the discovery of a million or more of Type Ia
supernovae per year of operation. This fantastic rate precludes
spectroscopic follow up on all but the smallest fraction.

Errors from spectroscopically determined redshifts ($\pm$ 0.001) are
typically much smaller than the errors associated with other
parameters used in constructing the Hubble diagram (e.g.,\ stretch and
magnitude) and are justifiably neglected. As future survey instruments
may discover thousands of Type Ia supernovae every night, a different
method for ascertaining redshifts certainly must be employed.
Photometric redshifts can be acquired much more expeditiously than
their spectroscopic counterparts (the latter typically requiring
numerous hours of dedicated time on a follow-up
telescope). Unfortunately, all such alternative methods introduce
non-negligible redshift error which must be accounted for in
subsequent analyses.

Fortuitously, both of the long-standing photometric redshift
techniques developed for galactic astronomy can be brought to bear on
assessing the photometric redshifts of Type Ia supernovae. As the
field of galactic redshift determination continues to mature, two
competing techniques have emerged. The first technique is based on
empirical fits derived from a training set having well established
redshifts \citep{1995AJ....110.2655C}.  Recently, an empirical
photometric estimator has been derived and applied to Type Ia
supernovae \citep{2007ApJ...654L.123W}. With this technique,
magnitudes derived from fluxes in the {\it grz} bands are linearly
combined to give a first-order redshift estimate. An improved redshift
estimate is obtained by adding a stretch-like correction involving a
ratio of the {\it i} band fluxes at peak and at 15 days in the
supernovae rest frame. The coefficients are derived from a SNLS
training set of 40 Type Ia supernovae. A dispersion ranging 0.031 to
0.050 is reported depending on the size of the training set. These
comparatively large numbers are attributed to the photometric quality
of the SNLS data and minimal statistics. In a follow up paper
\citep{2007MNRAS.382..377W} it was shown that a dispersion of 0.005
may be achievable from supernovae with S/N $>$ 25, albeit supernovae
that are temporally well sampled and free of obscuration from dust. A
second technique involves broad band photometry acquired over multiple
bands that is compared to predictions from galaxy spectral energy
distributions (SEDs) to determine the redshift
\citep{1986ApJ...303..154L} (and recently improved in
\citep{2000ASPC..200..392B}).  The photometric error from either
approach can be of order $\pm$ 0.1, which is generally unacceptable
for the study of individual galaxies of galaxy clusters though is less
problematic when performing statistical analyses of large data sets.

In this paper we adopt the latter approach to obtain photometric
redshifts from template fitting. The advantage of this approach is
that it obviates the need for an empirical model. However, a faithful
Type Ia template is required. The requirement of a faithful supernova
template is non-trivial since no two supernovae have identical
spectra, the spectra evolve with time and further exhibit variation in
luminosity. Further, Type Ia supernovae have no Hydrogen features in
their spectra and hence lack the Lyman-break used for determining
galactic redshifts. For template builders temporal evolution implies
more work, for photometric redshift estimation purposes it is a boon:
in essence each observing epoch permits an independent redshift
determination. Furthermore, all epochs for which there is photometric
data can be combined to jointly constrain the redshift.  The redshift
error which results when the template approach is applied to Type Ia
supernovae is the main objective of this paper.

In Section \ref{sec:meth} we describe the redshift error estimation
methodology based on an archetypal Type Ia supernovae template set. In
Section \ref{sec:res} we present results based solely on statistical
error, then extend them to include systematic error derived from a
distinct template set. From these combined statistical and systematic
errors we derive ($\Omega_{w}$,$\Omega_{w'}$) error ellipses based on
a realistic supernovae distribution and idealized cadence. We conclude
in Section \ref{sec:dis} with a discussion of the implications of our
work and discuss future extensions.

\section{Methodology} \label{sec:meth}

The template fitting method begins with a Type Ia supernova template
set. This set, typically generated in the supernovae rest frame, is
modified to mimic the effects of redshift (the modified template set
is henceforth referred to as the model set). Broad band filters are
applied to spectra at various epochs in the model set to generate
light curves. The light curve of a supernova whose redshift is to be
determined (the reference supernova) is then compared to the light
curves of the model set over a suitable redshift interval. The
redshift of the light curve from the model set that most closely
matches the light curve of the reference supernova (using a $\chi^2$
figure of merit) is assigned to the latter.

We adopt Nugent's Type Ia branch normal templates
\citep{2002PASP..114..803N} for our studies. These templates span the
temporal range from -16 day prior to +70 days after peak {\it V}
magnitude (for a total of 87 epochs. Fluxes for epochs -20 through -17
are set to zero in the Nugent template set). The wavelength coverage
is 1000 to 25000 \AA\ binned into 2400 10 \AA\ intervals. This set of
87 spectra constitutes the rest frame template set. There are four
operations that must be applied to this template set to generate the
model set (representing supernovae spectra at various redshifts in an
expanding universe). The first two each involve a reduction in the
flux by a factor of $(1+z)$, where $z$ runs from 0 to 1.0 in redshift
intervals of 0.001, to account for time dilation and cosmological
expansion (photon redshift) \citep{1972gcpa.book.....W}.  A third is a
time dilation factor of $(1+z)$ applied to each epoch
\citep{1996NuPhS..51..123G}. The fourth involves redshifting the
spectrum by $(1+z)$.

Our procedure then begins with Nugent's Type Ia branch normal template
set, where the interval of 1 day between successive epochs in the
supernova rest frame is replaced by a separation of $(1+z)$ days in
the frame of the observer. Two factors of $(1+z)$ factors were applied
to the LSST bandpasses\citep{2007ASPC..378..485I}, which incorporate
sky and filter transmission and CCD efficiency. Bandpass fluxes were
derived by convolving the $ugrizy$ LSST filter set with the Nugent
templates, after the templates were calibrated using a Johnson B band
filter. Specifically, the Johnson $B$ filter was convolved with the
Nugent template at time of peak epoch. The correction required to make
the resulting $B$ passband magnitude equal to -19.3 was then applied
to all subsequent data. The result of this exercise is the creation of
1001 light curve files, each containing the magnitudes for all 87
(time dilated) epochs in all six filters. This collection comprises
the model set.

A key aspect of our method involves the conversion of magnitudes into
colors. As is well known, the relationship between luminosity and flux
requires knowledge of the luminosity distance.  However, the
luminosity distance depends on the choice of cosmological models
luminosities. As will be made more explicit below, the conversion from
magnitudes to colors permits cancellation of the (unknown) luminosity
distance.

\begin{figure}[t] 
  \plotone{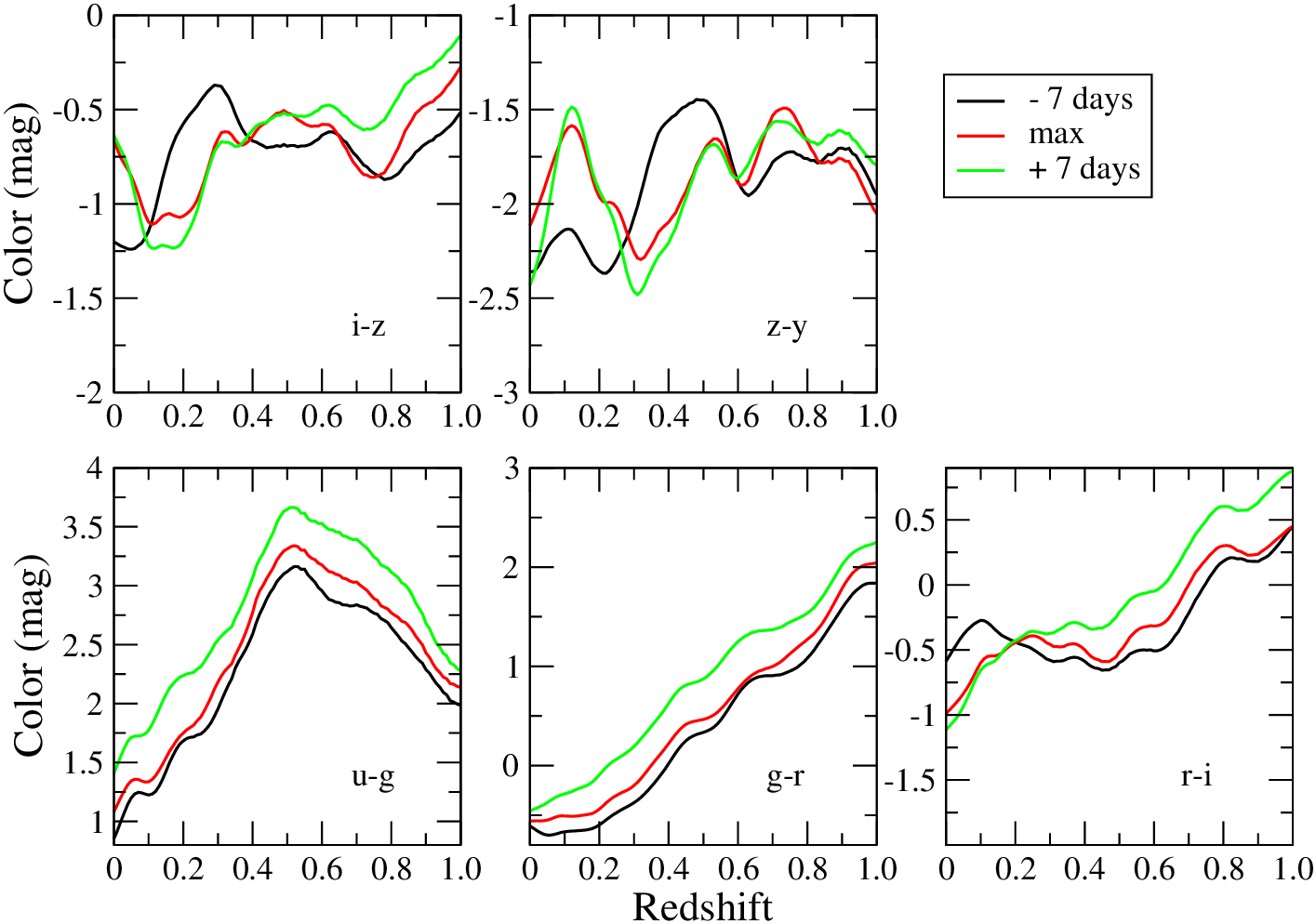}
  \caption{(lower left to upper right) LSST u-g, g-r, r-i, i-z and z-y
    colors as a function of redshift at 7 days before maximum (black),
    at maximum (red) and 7 days after maximum light (green) for the
    Nugent template set.} \label{fig:Fig6}
\end{figure}

Each panel in Figure \ref{fig:Fig6} shows the evolution of one the
five LSST colors obtained by applying the four corrections described
above to the LSST filter set convolved with Nugent's model template
set. Each color is shown as a function of redshift for three separate
epochs. Over the epoch range shown in this figure the $g-r$ color
exhibits nearly monotonic behavior and would be the most effective of
the five colors over the depicted redshift interval in estimating
redshifts.  The actual reconstruction efficiency will be considerably
more nuanced when multiple colors, multiple epochs and measurement
noise are involved, as we shall see below.
From these colors we now proceed to define a measure of goodness of
fit $chi^2(z)$ for the $u-g$ color as
\begin{equation}
  \chi^2(z) = \sum _i ^{epochs} ((u(z) - g(z)) - (u(z') - g(z')))_{i}^2,
\end{equation}\label{eq:chi}
where $u(z)$ and $g(z)$ represent magnitudes in the $u$ and $g$
bandpasses from the reference set at redshift $z$ at epoch $i$ and
$u(z')$ and $g(z')$ are defined similarly for the model set. The sum
is over all $i$ epochs.This procedure is generalized in an obvious
manner to include five colors ($u-g$, $g-r$, $r-i$, $i-z$ and $z-y$)
that can be constructed from the six LSST bandpasses.  Equation
\ref{eq:chi} makes explicit how luminosity distances cancel if one
chooses to work with color magnitudes. The $\chi^2$ expression in
Equation \ref{eq:chi} is actually a 1001 element array, with each
element of the array containing a $\chi^2$ value for a given
combination of $(z,z')$. The entire array is generated by fixing $z$
and varying $z'$ over the interval 0$\ < z' < \ $1 in increments of
0.001. The minimum of this array corresponds to the light curve at
{\it z'} from the model set that most closely matches the redshift of
the reference light curve at $z$. A perfect reconstruction results in
$z = z'$. Chi-squared values spanning the entire 1001 $\times$ 1001
$\{z,z'\}$ space are generated by choosing reference light curves over
the interval 0$\ < z < \ $1 in increments of 0.001.

The $\chi^2(z)$ value reflects the degree of overlap between the light
curve of a reference supernova located at redshift $z$ with respect to
the light curves of model supernovae located at $z'$. It bears
emphasizing that the reference and model sets are both drawn from the
same Nugent template set. The overlap that would result if spectral
and variability were included undoubtedly would be less pronounced.

\section{Results} \label{sec:res}

\begin{figure}[t]
  \plotone{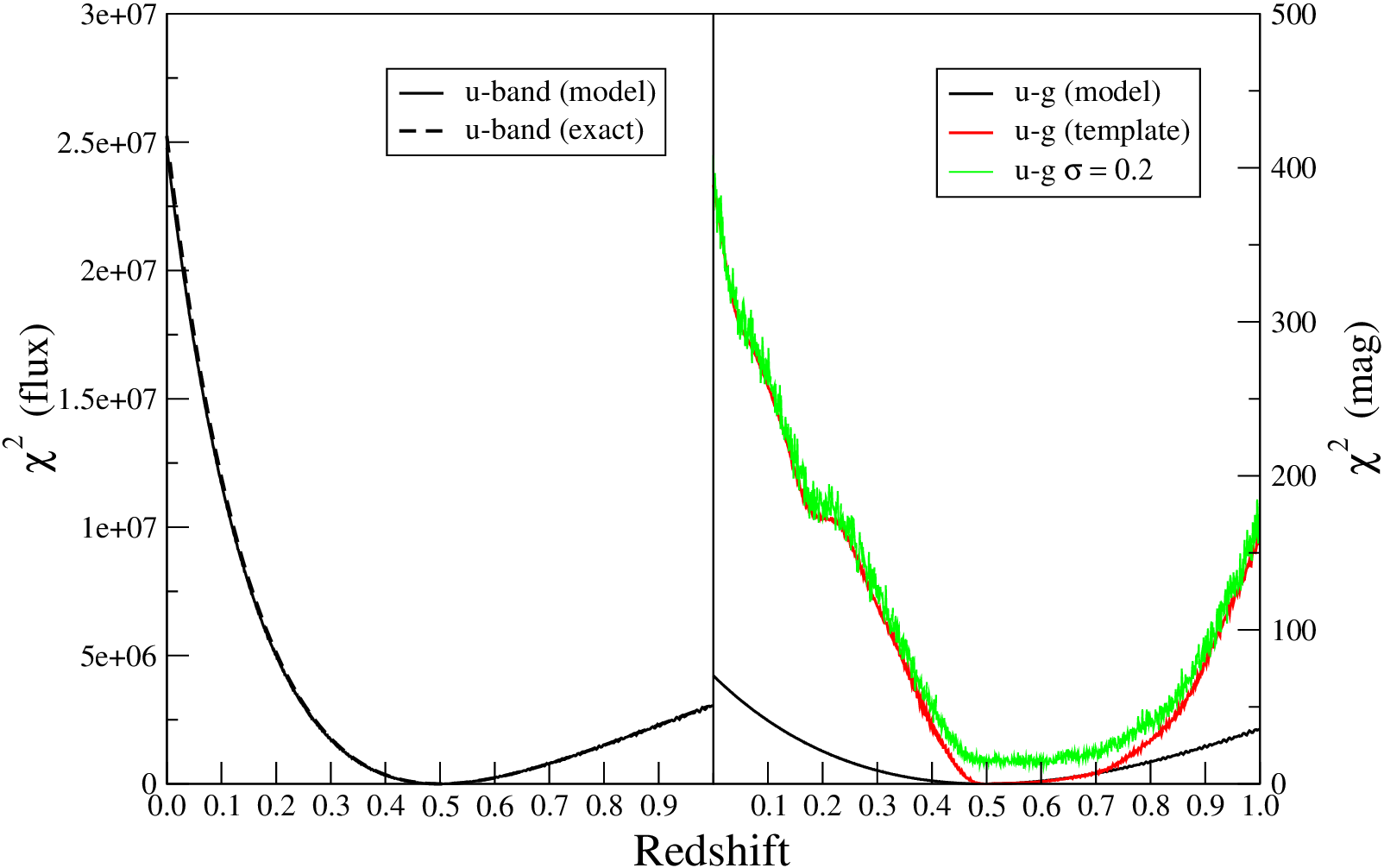}
  \caption{(Left) $\chi^2$ values (in units of flux) as function of
    redshift for a simplified analytical model (dashed) versus those
    derived from the numerical approach described in the text (solid)
    for a reference supernova at $z$ = 0.5. Small differences between
    the two approaches are attributed to binning of the latter, and
    the absence of noise the reconstructed redshift is 0.5 with either
    approach. (Right) $\chi^2$ values (in units of magnitudes) but for
    the LSST derived $u-g$ color with no noise (red), and with the
    addition of noise drawn from a normal distribution with $\sigma$ =
    0.05 (green). For comparison purposes, the result of the toy model
    is again shown, but now for the $u-g$ color (black, also in units
    of magnitudes). It is noteworthy that $\chi^2$ is much steeper for
    the template than the toy model. } \label{fig:Fig1}
\end{figure}

Though the methodology described above is straightforward, it is
instructive to explore a highly simplified model for the purposes of
comparison with results from using the template spectra. An analytical
expression for $\chi^2$ at an arbitrary redshift can be written down
if filter transmission coefficients and spectral fluxes (at all
epochs) are constant.  The left panel in Figure \ref{fig:Fig1}
compares the results from an analytic model (dashed) with those from
the methodology described in Section \ref{sec:meth} (solid) for a
reference supernova at a redshift 0.5. In both the analytical and
numerical examples the LSST $u$ bandpass is used, however, the
transmission coefficients have all been set to 1.0 over the rest frame
interval 3190 $ < \lambda < $ 4120 (those wavelength bins in the rest
frame $u$ filter whose transmission coefficient exceeds 0.005).  The
redshifted filter is convolved with a toy spectrum whose rest frame
fluxes are also set to 1.0 over all wavelengths and all epochs. Both
the analytical and numerical examples are described by Equation
\ref{eq:chi}, but here with $g(z) = g(z') = 0$. The close agreement
between them benchmarks the numerical approach.

The black curve in the right hand panel of this same figure is the
$\chi^2$ for the $u-g$ color as described by Equation \ref{eq:chi},
using the $u$ filter and toy spectrum as described above, and with a
rest frame $g$ filter whose transmission coefficients are 1.0 over the
region 3870 $ < \lambda < $ 5610. The featureless toy model produces a
$\chi^2$ curve whose values are non-zero only because the $u$ and $g$
bandpasses have different widths and are centered on different
wavelengths.  For comparison, the $\chi^2$ values for the $u-g$ color
derived from the Nugent templates and LSST bandpasses are also
displayed (red). Note that the $\chi^2$ curve derived from the
templates is much steeper than the toy model and further displays
structure which can be traced back, in principle, to various spectral
features. In the absence of noise the reconstructed redshift
corresponding to the $\chi^2$ minimum coincides with the reference
redshift of 0.5.  A more realistic example includes measurement
noise. The green curve in the right panel of Figure \ref{fig:Fig1} are
the $\chi^2$ values for the $ug$ color derived from the templates
where noise has been added to both the reference and model light
curves drawn from a normal distribution with standard deviation of
0.05. As this curve illustrates, noise obscures the location of the
true minimum in two ways: it increases bin-to-bin jitter and broadens
the minimum. Both effects increase the probability that the location
of the true minimum (located at $z'$) will differ from the
reconstructed minimum (located at $z$).

\begin{figure}[t]
  \plotone{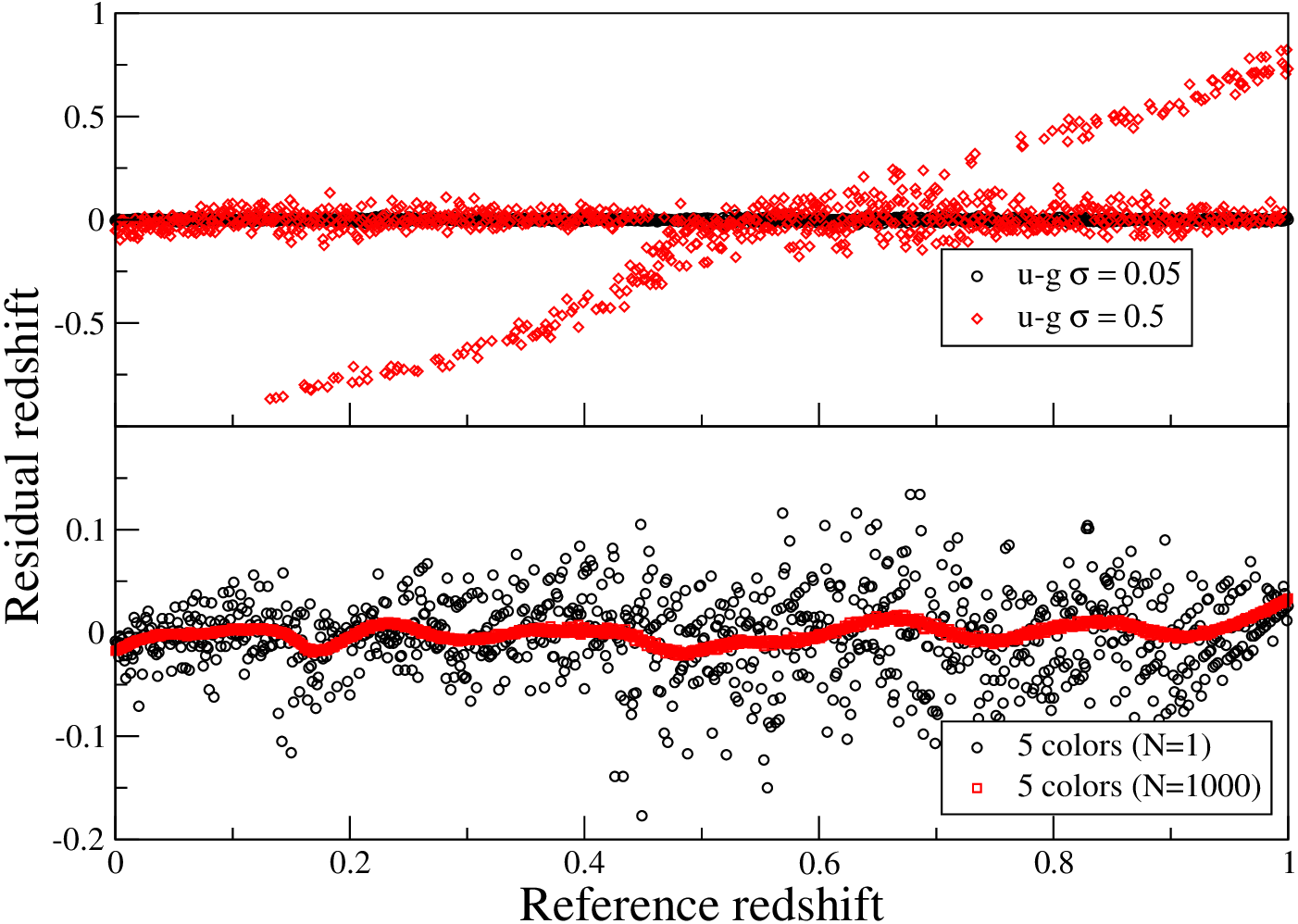}
  \caption{(Top) Residual redshifts as defined in the text for a
    single realization of noise drawn from a normal distribution with
    $\sigma$ = 0.05 and 0.5 computed for the LSST $u - g$ color
    assuming a nightly observing cadence.  (Bottom) Same, but for a
    single realization (circles) and the median of 1000 realizations
    (squares) of noise drawn from normal distribution with $\sigma$ =
    0.5 for all five LSST colors.} \label{fig:Fig2}
\end{figure}

\begin{figure}[t]
  \plotone{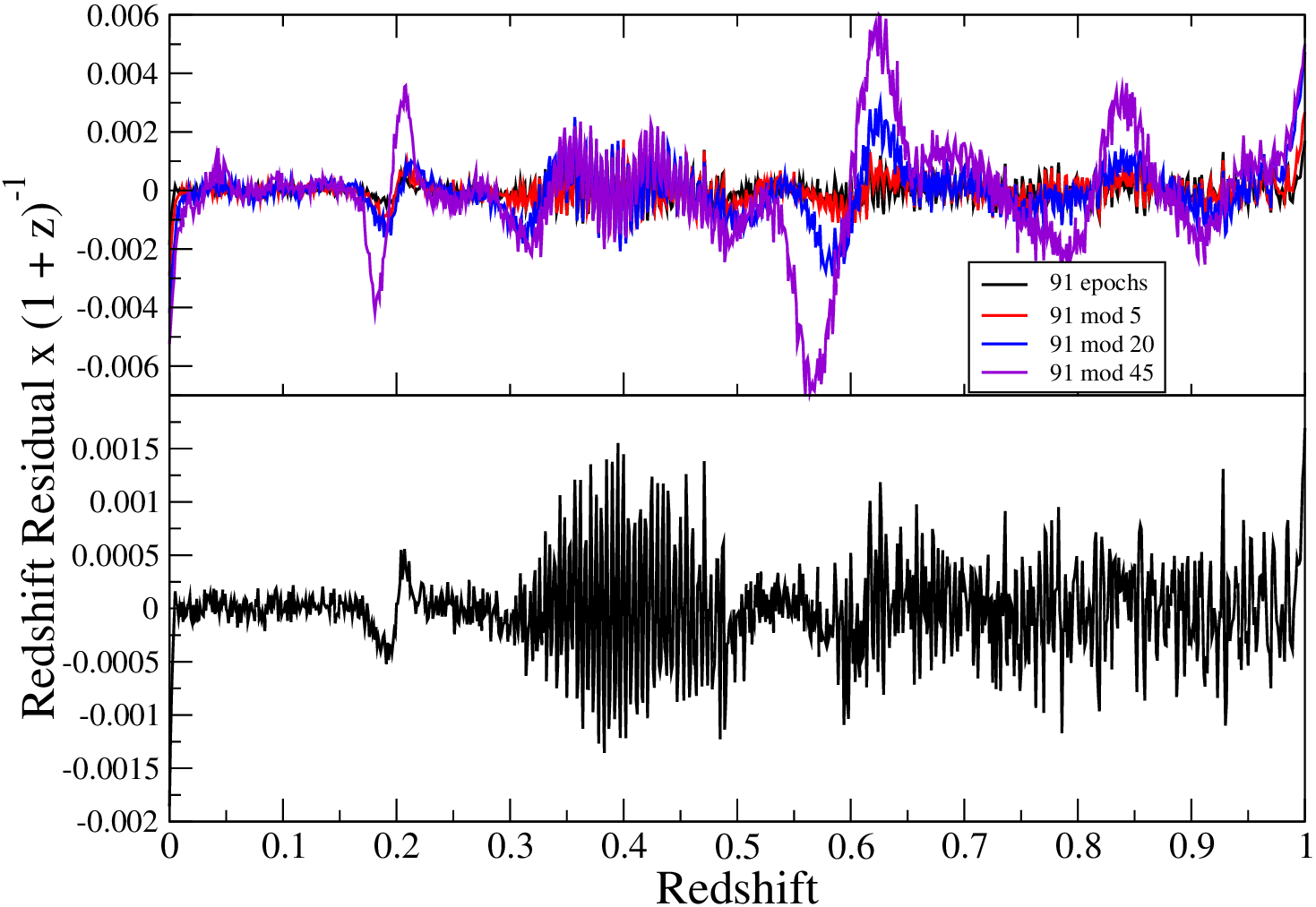}
  \caption{(Bottom) The median of the residual redshifts of 1000
    realizations drawn from a normal distribution with $\sigma$ = 0.05
    for all five LSST colors and assuming a nightly observing
    cadence. (Top) Same, but assuming an observing cadence of every
    5$^{th}$ night (red), 20$^{th}$ night (green) and 45$^{th}$ night
    (blue) night. For reference the curve from the bottom panel is
    also included (black).} \label{fig:Fig3}
\end{figure}

The solid curve in the right panel of Figure \ref{fig:Fig1} is a
single realization of Gaussian noise. To get a statistical measure of
the effect that noise has on obscuring the location of the true
minimum, Monte Carlo simulations consisting of 1000 random instances
of photometric measurement noise were done for each redshift
interval. From these runs the median redshift value was chosen and
compared to the redshift of the reference light curve.  The top panel
in Figure \ref{fig:Fig2} contains two curves, each a plot of the
residual redshift versus the redshift of the reference light curve
using LSST $u$ and $g$ bandpass filters. The residual redshift is
defined here as $(z_{rec} - z)/z$, where $z_{rec}$ is the
reconstructed redshift based on the above procedure and $z$ is the
redshift of the reference light curve. The sole difference between the
curves in the upper panel of Figure \ref{fig:Fig2} is the amount of
Gaussian noise added to the reference and model light curves. It is
seen that reconstruction of the redshift is quite faithful for the
addition of Gaussian noise sampled from a normal distribution with a
$\sigma$ of 0.05, but erratic behavior is observed with a $\sigma$ of
0.5. The bifurcation one sees in this panel for large photometric
errors is due to non-monotonic nature of color versus redshift curve
in Figure \ref{fig:Fig1} ($u-g$ panel).  Since for z $>$ 0.2
(depending on epoch) the $u-g$ color function is double-valued, the
$\chi^2$ minimization procedure alternates between two possible
redshift solutions, giving rise to the aberrant behavior of the red
curve in Figure \ref{fig:Fig3}.  By this same logic from Figure
\ref{fig:Fig1} one can predict more rational behavior for $z \sim$
0.5, which is in fact manifested in Figure \ref{fig:Fig3}.

As the lower panel demonstrates, the erratic behavior in the top panel
can be modified using all five colors combinations of the LSST filter
set. Though considerable bin-to-bin scatter is observed in the
reconstructed redshift from only a single realization of Gaussian
noise (circles), a 1000 realizations generates the smoothly varying
curve (squares) in this same panel. Notably, the data are well-behaved
with no catastrophic outliers even with the addition of considerable
noise.  As noted above, the observed undulations ultimately can be
traced to template spectral features passing through the filter set.
The lower panel in Figure \ref{fig:Fig3} shows the residual redshifts
using all five bandpass colors at every epoch, but now with Gaussian
noise drawn from a normal distribution with a $\sigma$ of 0.05. In
this plot the residual redshifts are divided a factor of $(1 + z)$.
For this highly idealized cadence the standard deviation of the
redshift residuals is 4.2$\cdot$10$^{-4}$ over the redshift interval 0
$\leq z \leq$ 1.0. 

\begin{figure}[t]
  \plotone{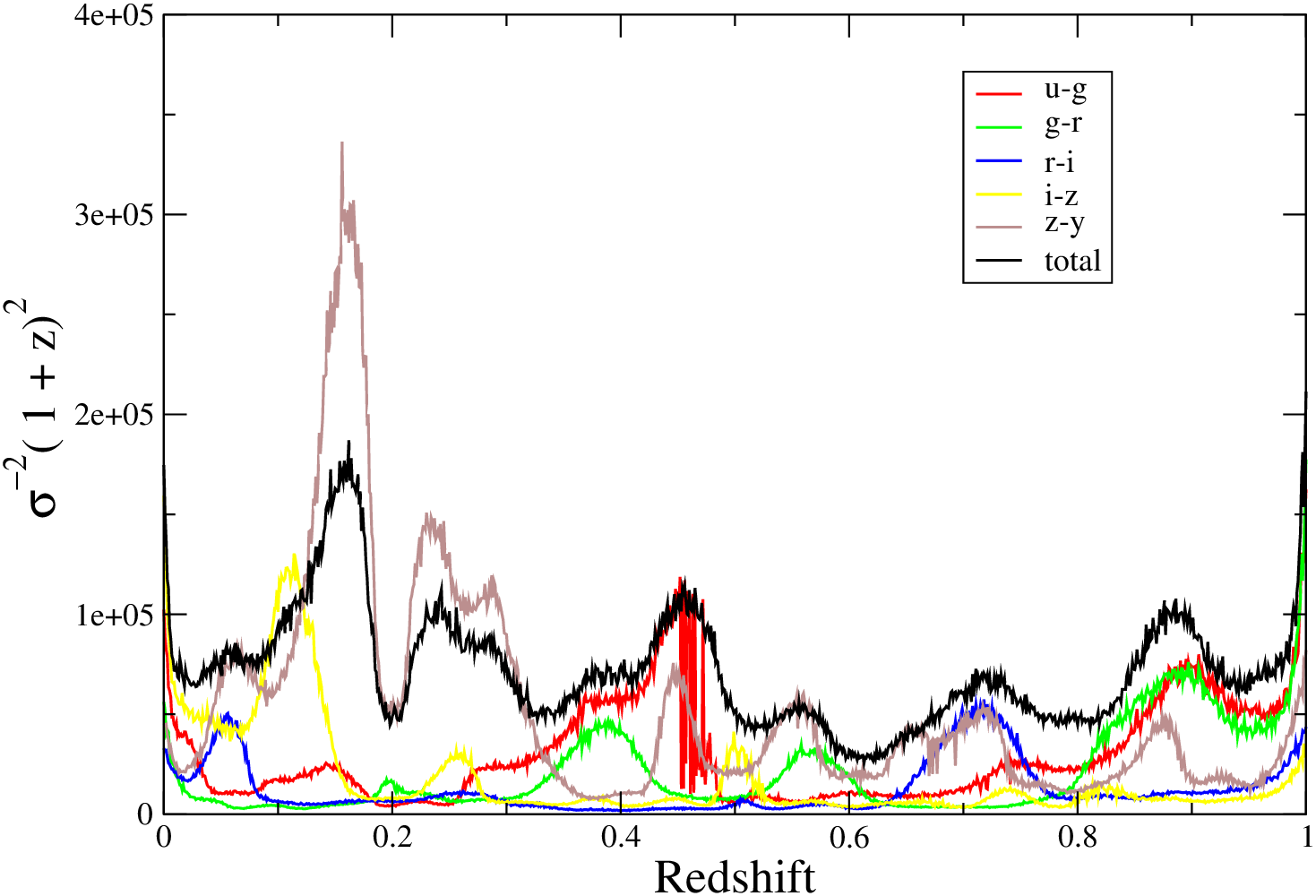}
  \caption{The inverse square of the errors of the redshift residuals
    from 1000 realizations of noise from a normal distribution with
    $\sigma$ = 0.05 and a observing cadence of every 5$^{th}$ night in
    each of the five LSST colors and the total inverse square error
    (black).}\label{fig:Fig4}
\end{figure}

The redshift residuals presented so far have been constructed using
all 87 epochs. It is certainly unrealistic for a survey instrument to
sample any given supernovae on a nightly basis. The upper panel in
this same figure shows residual redshift results for successively
sparser sampling intervals. As a benchmark the black curve from the
lower panel is included, along with the residuals for cadences
representing observations using all six LSST filters
 every 5$^{th}$
(red), 20$^{th}$ (green) and 45$^{th}$ (blue) night. The corresponding
standard deviations are 5.2$\cdot$10$^{-4}$, 9.2$\cdot$10$^{-4}$ and
1.8$\cdot$10$^{-3}$, respectively. Clearly, there are redshift
intervals for which even an extremely sparse cadence may provide an
accurate photometric redshift estimate. In any case, the data can be
corrected for the bias (departure from perfect reconstruction) seen in
these figures. By contrast, {\it errors} associated with the redshift
residuals cannot be compensated for. Figure \ref{fig:Fig4} is a plot
of the inverse square of the residual redshift errors for each of the
five LSST colors assuming an observing cadence of every fifth
epoch. In this plot larger values correspond to smaller errors and
illustrates the "effectiveness" of each color for getting accurate
redshift in a given redshift range. This same plot also shows the
total errors when all five colors are used (black curve). Note that
even though $\sigma^{-2}$ is an additive quantity, the sum from each
of the individual colors does not equal that from the combined five
colors since the colors are not linearly independent.  The
effectiveness of the $z-y$ color is particularly
striking. Specifically, a conceivable strategy could be to restrict
photometry to the $z$ and $y$ filters over the redshift interval 0.15
$ < z < $ 0.3 should the templates resemble actual supernovae spectra
in these bandpasses.
As an optimal supernovae cadence for LSST has yet to be adopted,
subsequent analyses are based on the assumption that a particular
supernova will be visited in each of the six filters every fifth
night.

In addition to the statistical error arising from Poisson noise and
imperfect sampling, there may be a plethora of systematic errors that
will further degrade photometric redshift accuracy. One systematic
error than is relevant here concerns choice of templates. To generate
binned systematic errors we chose a second model template set
constructed by \citep{2007ApJ...663.1187H}. This set has identical
spectral coverage as the Nugent set, but epoch coverage is extended to
+85 days (for this purpose epochs beyond day 70 are ignored). We
derive systematic errors by again computing the standard deviation in
the redshift residuals for 1000 trails sampling from a normal
distribution with $\sigma = 0.05$ and also very fifth night, but now
using this second model set. The standard deviations that result are
taken to be systematic error due to imperfect knowledge of the
template set.

\section{Cosmological Parameters} \label{sec:cp}
\begin{figure}[]
  \plotone{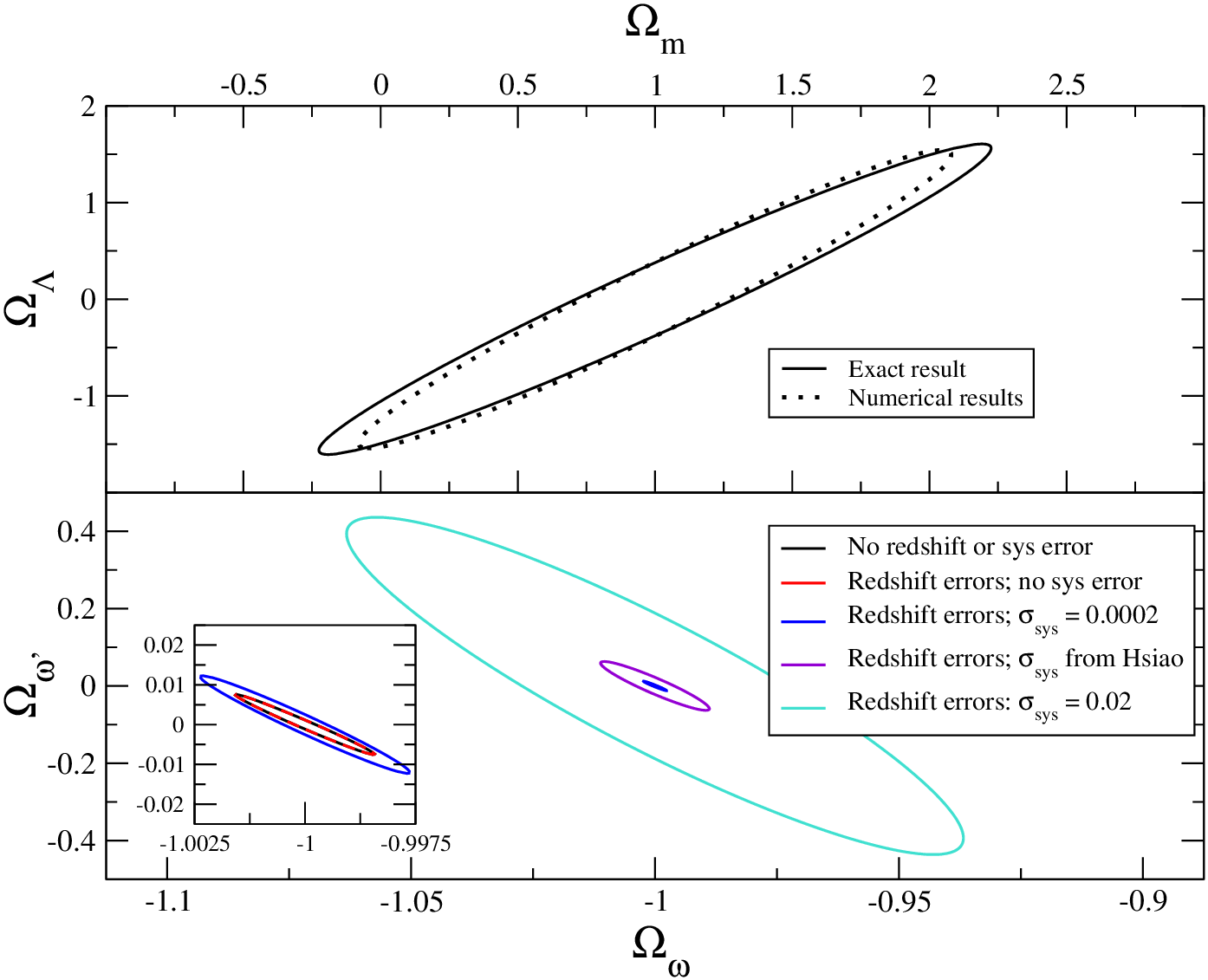}
  \caption{(Top) Error ellipses in the
    \{$\Omega_{M}$,$\Omega_{\Lambda}$\} plane from Fisher matrices
    derived from an analytical (solid) and discrete model (dashed) of
    the Gaussian supernovae distribution as described in the text. In
    both cases a flat, matter-filled universe ($\Omega_{M}$ = 1,
    $\Omega_{\Lambda}$ = 0) was assumed. (Bottom) Same, but now in the
    \{$\Omega_w$, $\Omega_{w'}$\} plane and assuming the standard
    cosmology ($\Omega_{M}$ = 0.3, $\Omega_{\Lambda}$ =
    0.7). A series of five error ellipses are shown, with the blue
    ellipse representing the addition of a generic systematic error
    with a $\sigma_{sys}$ of 0.0002. The violet ellipse represents the
    addition of systematic error derived from utilizing a distinct
    model template set (see text). The aquamarine ellipse represents a
    generic systematic error with a $\sigma_{sys}$ of 0.02 as in
    \citep{2004ApJ...615..595H, 2008ApJ...675L...1Z}. In all three
    cases the systematic error is added in quadrature to the redshift
    errors described by the plot in the lower panel of Figure
    \ref{fig:Fig3} for the supernovae distribution described in
    \citep{2008ApJ...675L...1Z}.  The inset shows on an expanded scale
    the three smallest error ellipses from the main panel representing
    pure photometry error (black), photometry and redshift errors
    (red). The blue ellipse is included in the inset for
    comparison. }\label{fig:Fig5}
\end{figure}

Ultimately, one wishes to transform the redshift errors arising from
the imperfect photometric redshift reconstruction onto the accuracy
with which one can determine cosmological parameters. A common
approach for propagating redshift errors from data, or estimated
errors from future data sets, is via the Fisher matrix. The Fisher
matrix is a statistic which combines data errors with model
sensitivities and whose inverse is the covariance matrix. Several
techniques have been proposed for propagating redshift error using the
Fisher matrix formalism; here we employ the method advanced by
\citep{2004ApJ...615..595H}. In their approach an $N\times N$ Fisher
matrix (representing $N$ cosmological parameters) is expanded by $M$
rows and columns representing the $M$ supernovae whose redshift errors
are to be marginalized over. The off-diagonal terms in these $M$ rows
and columns contain cross products of the derivatives of the distance
modulus $\mu$ with respect to redshift times the derivatives of the
distance modulus with respect to $\Omega_{M}$ or $\Omega_{\Lambda}$,
e.g., $\partial \mu/ \partial z \cdot \partial \mu/ \partial
\Omega_{M}$. The $M$ diagonal terms contain inner products of the form
$\partial \mu/ \partial z \cdot \partial \mu/ \partial z$ and all
$M\times M$ terms are evaluated at the (binned) supernova redshift and
weighted by the binned errors. When constructing the quantity
$\partial \mu/ \partial z$ the definition of $\mu$ is expanded to
include stretch (but not derivatives of the $K$-corrections as was
required in \citep{2004ApJ...615..595H} as we are working with
magnitudes derived directly from the spectra, not the light curves).
As a cross check this discrete method was compared with an analytical
Gaussian distribution proposed by \citep{1998astro.ph..4168T}. The
upper panel in Figure \ref{fig:Fig5} compares the exact result with the
binned approach for a Gaussian distribution of 100 supernovae at an
average redshift of $z$ of 0.55, with a standard deviation of 0.2 and
each with a magnitude error of 0.5 for a flat, matter-filled universe
($\Omega_{M}$=1, $\Omega_{\Lambda}$=0).
The variation between the two approaches is ascribed to the discrete
binning (16 bins) that all but disappears as the number of bins
approaches 100 (at the expense of computational resources required to
invert a 102 $\times$ 102 matrix).  

A survey instrument like LSST with an \'etendue of 220 m$^2$deg$^2$
will cover the entire sky every few nights in each of its six
filters. Though its primary cadence may not be optimal for detection
and follow up of Type Ia supernovae, LSST is nevertheless expected to
discover several million Type Ia supernova in the course of its 10
year mission. This sheer number of supernovae will allow unprecedented
examination of systematics related to reddening, host contamination,
evolution and other effects that interfere with a precise
determination of the distance modulus. Lacking a formal supernovae
cadence for LSST we adopt the estimated number of supernovae per
steradian per unit redshift per year in the observer frame as derived
by \citep{2008ApJ...675L...1Z}.

The ellipses in the lower panel in Figure \ref{fig:Fig5} are derived
from combined photometry, photometric redshift and systematic
errors. Photometry errors ($\sigma_p$ = 0.17) and photometric redshift
errors associated with the orange curve from the upper panel in Figure
\ref{fig:Fig4} ($\sigma$ = 0.05; every 5$^{th}$ epoch) are first
reduced by the number of supernovae in each of the rebinned redshift
bins (16 equally spaced bins spanning 0 $< z < 1.6$; weights for $z >$
1.0 were made large to effectively limit the analysis to $0 < z <
1.0$). Systematic error derived in the manner described above was
likewise rebinned and added in quadrature to the photometry and
photometric redshift error. A 2 $\times$ 2 Fisher matrix representing
the cosmological parameters $\Omega_w$ and $\Omega_w'$ was expanded to
an 18 $\times$ 18 matrix to accommodate these binned errors. This
matrix was inverted and marginalized along the 16 rows and columns
representing the binned errors. The resultant 2 $\times$ 2 inverse
matrix was again inverted to generate the purple ellipse in the lower
panel of Figure \ref{fig:Fig5}. Several related error ellipses are
shown in this same panel for comparison purposes. Photometry error
alone produces the black ellipse - addition in quadrature with
photometric redshift errors produce an ellipse (red) that
is indistinguishable from the black ellipse (see inset). This
behavior, also noted by \citep{2004ApJ...615..595H}, is due to the
particular choice of photometry error and will evolve as photometry
errors become more stringent. The blue ellipse demonstrates that
systematic errors have no tangible effect until it reaches a level that
coincides with the level of systematic error derived from the Hsiao
template set ($\sigma_{sys} \sim$ 0.003). The largest ellipse
(aquamarine) assumes a $\sigma_{sys}$ of 0.02 in keeping with
assumptions made by \citep{2004ApJ...615..595H, 2008ApJ...675L...1Z}.

\section{Discussion} \label{sec:dis}

A new era of large aperture ground-based survey telescopes is driving
the need for an alternative to costly spectroscopic follow up to
obtain Type Ia supernovae redshifts. In this paper we apply a
technique that utilizes templates to derive these redshifts
photometrically based on a $\chi^2$ minimization procedure. Though
applied to proposed LSST bandpasses, this approach is broadly
applicable to any instrument acquiring photometric data.  Limited
Poisson statistics, observational and instrumental noise are modeled
by the addition of Gaussian noise. Further, we approximate a realistic
supernovae campaign cadence by restricting the number of epochs used
to construct our $\chi^2$ figure of merit whose minimum is assigned to
the redshift of the reference supernova.  Multiple realizations of
these simulations generate the statistics from which we extract the
residuals and their standard deviations for each redshift
bin. Ultimately, of course, we are interested in the implications for
cosmology with redshifts derived in this manner. To proceed, we
overlay a realistic redshift distribution broken into observing
bins. These redshift errors are then placed in a matrix along with the
weights of the cosmological parameters of interest, with a common
multiplier reflecting photometry errors. Both the redshift errors and
photometry errors are weighted by the number of supernovae per bin;
binned systematic errors derived from a distinct model set are added in
quadrature but are not weighted by the number of supernovae per
bin. The resulting Fisher matrix generates error ellipses in the
$\Omega_{\Lambda}$ and $\Omega_{M}$ and $\Omega_w$ and $\Omega_{w'}$
planes.

Though considerable work has been done to ascertain the feasibility of
weak-lensing measurements from ground-based instruments
\citep{2005ApJ...632L...5W,2006JCAP...02..001J,2007ApJ...659...69A},
comparatively little has been done regarding Type Ia supernovae. This
work is the first to put all the major elements together to ascertain
the relevance of supernovae cosmology from ground-based
instruments. In particular, it establishes a minimal redshift error
that can be expected from photometric redshift determination.  The
error ellipse derived in Figure \ref{fig:Fig5} from plausible
photometric redshift and systematic errors is considerably more
constrained than would be the case if one assumed constant photometric
redshift and systematic errors of magnitude 0.02 that have appeared in
the literature \citep{2004ApJ...615..595H, 2008ApJ...675L...1Z}. A
corollary is that should systematic error ultimately be controllable
to $\sim$ 0.003, then photometric redshifts would be very effective at
discriminating between many competing cosmologies.

A more realistic observing cadence, Type Ia supernovae distribution
and noise models are readily incorporated into this framework. Indeed,
though the scope of this work has been strictly limited to template
models, it is readily modified to actual supernovae spectra. Progress
is being made extending this work to include simulated cadences and
actual Type Ia supernovae spectra and work continues on improved noise
models. Extensions of this work must further address covariances among
the techniques used for typing, epoch and redshift assignments.

We conclude with a final observation. At present this method is
limited to stretch $\sim$ 1 supernovae since, at present, there is no
firm consensus as to the underlying mechanism for stretch. However,
correlations between various spectral features and stretch are well
established \citep{1995ApJ...455L.147N, WWA} and with these
correlations in hand we have embarked on the construction of Type Ia
supernovae templates with stretch different from one.  With these
templates in hand the above technique could be extended by simply the
addition of a loop over templates. One advantage to templates
constructed in this manner is that the variability would not
necessarily be restricted to that observed in actual Type Ia
supernovae \citep{2007A&A...466...11G}.

%% Putting eqnarrays or equations inside the mathletters environment
%% groups the enclosed equations by letter. For instance, the eqnarray
%% below, instead of being numbered, say, (4) and (5), would be
%% numbered (4a) and (4b).  LaTeX the paper and look at the output to
%% see the results.

%% separate it off from the body of the text using the
%% \acknowledgments command.

%% Included in this acknowledgments section are examples of the AASTeX
%% hypertext markup commands. Use \url without the optional [HREF]
%% argument when you want to print the url directly in the
%% text. Otherwise, use either \url or \anchor, with the HREF as the
%% first argument and the text to be printed in the second.

\acknowledgments

We wish to acknowledge Dragan Huterer for help in implementing his
Fisher matrix analysis.

\clearpage


\begin{thebibliography}{}
\bibitem[Riess et al.(1998)]{1998AJ....116.1009R} Riess, A.~G., et
  al.\ 1998, \aj, 116, 1009
 
\bibitem[Perlmutter et al.(1999)]{1999ApJ...517..565P} Perlmutter, S.,
  et al.\ 1999, \apj, 517, 565

\bibitem[Astier et al.(2006)]{2006AA...447...31A} Astier, P., et al.\
  2006, \aap, 447, 31

\bibitem[Wood-Vasey et al.(2007)]{2007ApJ...666..694W} Wood-Vasey,
  W.~M., et al.\ 2007, \apj, 666, 694

\bibitem[Connolly et al.(1995)]{1995AJ....110.2655C} Connolly, A.~J.,
  Csabai, I., Szalay, A.~S., Koo, D.~C., Kron, R.~G., \& Munn, J.~A.\
  1995, \aj, 110, 2655

\bibitem[Loh \& Spillar(1986)]{1986ApJ...303..154L} Loh, E.~D., \&
  Spillar, E.~J.\ 1986, \apj, 303, 154

\bibitem[Bolzonella \& Pell{\'o}(2000)]{2000ASPC..200..392B}
  Bolzonella, M., \& Pell{\'o}, R.\ 2000, Clustering at High Redshift,
  200, 392

\bibitem[Wang(2007)]{2007ApJ...654L.123W} Wang, Y.\ 2007, \apjl, 654,
  L123

\bibitem[Wang et al.(2007)]{2007MNRAS.382..377W} Wang, Y., Narayan,
  G., \& Wood-Vasey, M.\ 2007, \mnras, 382, 377

\bibitem[Nugent et al.(2002)]{2002PASP..114..803N} Nugent, P., Kim,
  A., \& Perlmutter, S.\ 2002, \pasp, 114, 803

\bibitem
  www.phy.bnl.gov/\~partsem/fy07/BNL\_Wood-Vasey\_20070719.key/droppedImage-8.pdf

\bibitem[Weinberg(1972)]{1972gcpa.book.....W} Weinberg, S.\ 1972,
  Gravitation and Cosmology: Principles and Applications of the
  General Theory of Relativity, by Steven Weinberg, pp.~688.~ISBN
  0-471-92567-5.~Wiley-VCH, July 1972

\bibitem[Goldhaber et al.(1996)]{1996NuPhS..51..123G} Goldhaber, G.,
  et al.\ 1996, Nuclear Physics B Proceedings Supplements, Vol.~51,
  51, 123

\bibitem[Ivezi{\'c}(2007)]{2007ASPC..378..485I} Ivezi{\'c}, {\v Z}.\ 2007, 
Why Galaxies Care About AGB Stars: Their Importance as Actors and Probes, 
378, 485

\bibitem[Hsiao et al.(2007)]{2007ApJ...663.1187H} Hsiao, E.~Y.,
  Conley, A., Howell, D.~A., Sullivan, M., Pritchet, C.~J., Carlberg,
  R.~G., Nugent, P.~E., \& Phillips, M.~M.\ 2007, \apj, 663, 1187

\bibitem[Huterer et al.(2004)]{2004ApJ...615..595H} Huterer, D., Kim,
  A., Krauss, L.~M., \& Broderick, T.\ 2004, \apj, 615, 595

\bibitem[Tegmark et al.(1998)]{1998astro.ph..4168T} Tegmark, M.,
  Eisenstein, D.~J., \& Hu, W.\ 1998, arXiv:astro-ph/9804168

\bibitem[Zhan et al.(2008)]{2008ApJ...675L...1Z} Zhan, H., Wang, L.,
  Pinto, P., \& Tyson, J.~A.\ 2008, \apjl, 675, L1

%\bibitem[Albrecht et al.(2006)]{2006astro.ph..9591A} Albrecht, A., et
%  al.\ 2006, arXiv:astro-ph/0609591

\bibitem[Nugent et al.(1995)]{1995ApJ...455L.147N} Nugent, P.,
  Phillips, M., Baron, E., Branch, D., \& Hauschildt, P.\ 1995, \apjl,
  455, L147

\bibitem[Wittman(2005)]{2005ApJ...632L...5W} Wittman, D.\ 2005, \apjl, 632, 
L5 

\bibitem[Jain et al.(2006)]{2006JCAP...02..001J} Jain, B., Jarvis, M., 
\& Bernstein, G.\ 2006, Journal of Cosmology and Astro-Particle Physics, 2, 1 

\bibitem[Asztalos et al.(2007)]{2007ApJ...659...69A} Asztalos, S., de 
Vries, W.~H., Rosenberg, L.~J., Treadway, T., Burke, D., Claver, C., Saha, 
A., \& Puxley, P.\ 2007, \apj, 659, 69 

\bibitem[Wagers et al.(2009)]{WWA} Wagers, A., Wang, L., Asztalos, S.,
  Submitted to \apj

\bibitem[Guy et 
al.(2007)]{2007A&A...466...11G} Guy, J., et al.\ 2007, \aap, 466, 11 

\end{thebibliography}
\end{document}